%% file: Rapp_signal_distortions_codir_Raman_arxiv.tex
\documentclass[10pt]{article}

\usepackage{cite}
\usepackage{a4wide}
\usepackage{amsmath,amssymb}
\usepackage[colorlinks=false,bookmarks=false,citecolor=false,urlcolor=black]{hyperref}
\usepackage{graphicx}
\usepackage{tikz}
\usepackage{url}
\usepackage{caption}
\usepackage{svg}
\usepackage{color}
\usepackage{colortbl}
\usepackage{soul}
\usepackage{xfrac}

\DeclareGraphicsExtensions{.eps}
\graphicspath{{figs/}}

\newlength{\colwidth}
\newlength{\figwidth}
\newlength{\capwidth}
\newcommand{\graficeps}[4]{
\begin{figure}[#4]
\centering
\includegraphics[width=#2]{#1.eps}\\
\caption[#3]{\label{#1}\parbox[t]{\capwidth}{#3}}
\end{figure}
}

\definecolor{red}{rgb}{1,0,0}
\definecolor{green}{rgb}{0,0.7,0}
\definecolor{blue}{rgb}{0,0,1}
\definecolor{gray}{cmyk}{0, 0, 0, 0.7}
\definecolor{lightgray}{cmyk}{0, 0, 0, 0.117}
\definecolor{black}{cmyk}{0, 0, 0, 1.0}
\definecolor{lightyellow}{rgb}{1,1,0.8}
\definecolor{grey}{rgb}{0.8,0.8,0.8}
\definecolor{lightblue}{rgb}{0.8,0.8,1.0}
\definecolor{lightgreen}{rgb}{0.8,1.0,0.8}
\definecolor{lightyellow}{rgb}{1.0,1.0,0.8}
\definecolor{lightorange}{rgb}{1.0,0.9,0.8}
\definecolor{lightred}{rgb}{1.0,0.8,0.8}
\definecolor{lightpink}{rgb}{1.0,0.8,0.95}

\newcommand{\efield}{\ensuremath{\tilde{E}}}

\newcommand{\signaldet}{\ensuremath{\efield_{\rm signal,det}}}
\newcommand{\signalRx}{\ensuremath{\efield_{\rm signal,Rx}}}
\newcommand{\signalTx}{\ensuremath{\efield_{\rm signal,Tx}}}
\newcommand{\noise}{\ensuremath{\bf\tilde{n}}}

\newcommand{\phasecommon}{\ensuremath{\bf\Delta\varphi}}
\newcommand{\phasecommonest}{\ensuremath{\bf\Delta\varphi_{\rm est}}}

\newcommand{\phaseactual}{\ensuremath{\bf\phi}}
\newcommand{\weightfactor}{\ensuremath{{\cal{W}}}}
\newcommand{\probfactor}{\ensuremath{{\alpha_{\rm prob}}}}
\newcommand{\phaseconst}{\ensuremath{{\beta_1}}}
\newcommand{\dphaseconst}{\ensuremath{{\Delta\beta_1}}}
\newcommand{\alphaNp}{\ensuremath{{\alpha_{Np}}}}
\newcommand{\alphadB}{\ensuremath{{\alpha_{dB}}}}

\input{units}

\begin{document}
\title{Improving Performance of Higher--Order Codirectional Raman Amplifiers Using Phase--Modulated Signals --- Functional Principle}
\author{Lutz~Rapp
\thanks{L. Rapp is with ADVA SE, Märzenquelle 1--3, 98617 Meiningen, Germany (e--mail: lrapp@adva.com)}}
\markboth{Performance improvement with higher--order codirectional Raman pumping, \today}{Performance improvement with higher--order 
codirectional Raman pumping, \today}
\maketitle

\begin{abstract}
Existing unrepeatered submarine links are increasingly upgraded to the most advanced modulation format currently available for 
commercial applications. Quite often the use of third--order codirectional Raman amplifiers is necessary. Power fluctuations 
of the involved high--power pump induces phase shifts in phase modulated signals via the nonlinear Kerr effects. An approach 
for reducing the impact of this effect on system performance is explained in this document. 
\end{abstract}

\section{Introduction}
Upgrading existing fiber links with transponders operating at higher data rates typically requires to improve the optical signal--to--noise 
ratio (OSNR). Codirectional Raman amplification is an attractive technology to increase the performance of optical fiber links~\cite{Krummrich-ECOC-01}. 
The achievable performance improvement originates from a modified power profile in the transmission fiber that allows to achieve higher output powers 
at the output of the transmission fiber with comparable impact of nonlinear fiber effects. This finally leads to an improved optical signal--to--noise 
ratio (OSNR)~\cite{Murakami-ECOC-01}. With the trend to advanced modulation formats at higher data rates employing phase--shift keying, this technology 
is gaining significantly in importance in unrepeatered submarine links, but increasingly also in terrestrial systems. 

In many cases, using this kind of amplification scheme is the only remaining option for upgrading existing unrepeated submarine links by employing 
transponders with higher data rates. In the past, preferably counterdirectional Raman amplifiers~\cite{Evans-OFC-01} or even remote optically pumped 
amplifiers (ROPAs)~\cite{Lucero-OFC-09} have been installed in order to enable communication over long distances without intermediate active components 
requiring electrical power supply. With the introduction of advanced modulation formats, the OSNR at the receivers needs to be further increased due to 
the larger electrical bandwidth or decreased Euclidian distance between the symbols, since sophisticated forward error correction (FEC) codes and the 
benefits of coherent detection can only partly mitigate the requirement for better signal quality. 

Higher--order pumping schemes can provide a further noticeable improvement of the system performance when using counterdirectional Raman 
amplification~\cite{Papernyi-OFC-02}\cite{Faralli-OFC-05}\cite{Schneiders-JLT-06} by moving the gain from the end of the fiber span deeper into the 
fiber. In contrast, there is almost no benefit by applying this pumping scheme to already installed links using ROPAs since the required relocation 
of the ROPA cassette does not make economic sense~\cite{Pavlovic-SubOptic-13}. Commercial systems are available up to third--order, but results for 
schemes making use of sixth order pumping have already been published~\cite{Papernyi-OFC-05}. Till now, this technique has successfully been used in 
submarine links, but is also becoming attractive for terrestrial systems. 

The big advantage of codirectional Raman amplification is its compatibility with all the above mentioned techniques and that it can be installed in 
existing links in a cost--effective manner~\cite{Rapp-Suboptic-10}. Furthermore, higher orders are also applicable to codirectional 
pumping~\cite{Schulze-OFC-05}. However, a major drawback of higher--order pumping schemes is the decrease of the efficiency with increasing order. 
Therefore, the large pump powers required for third--order pumping are difficult to produce by multiplexing laser diodes. As a consequence, Raman 
fiber lasers (RFLs) are employed in most cases. However, these sources are known for their significantly higher relative intensity noise (RIN) as 
compared with laser diodes. Unfortunately, this gives rise to additional effects deteriorating system performance and achieved performance improvement 
is typically significantly smaller than predicated by theoretical considerations considering the modified power profile only. 

The transfer of relative intensity noise (RIN) from the pump to the signals via stimulated Raman scattering (SRS) has been investigated intensively 
and detailed mathematical models have been elaborated for first order~\cite{Fludger-JLT-01}\cite{Lakoba-JLT-04}\cite{Martinelli-PTL-05} and second 
order pumping schemes~\cite{Mermelstein-JLT-03}. These investigations have a strong focus on the frequency characteristics and deal with intensity 
modulated signals. Even techniques for suppressing this effect have been proposed~\cite{Mermelstein-PTL-03}. 

However, third--order codirectional Raman amplification also introduces random phase variations. These phase variations also go back to intensity 
variations of the pump lasers, but the impact on signal phase is caused by cross--phase modulation (XPM)~\cite{Agrawal-NLFO-07} and cross 
polarization modulation (XPolM)~\cite{Louchet-ICTON-11}. Due to the large bandwidth of the Kerr effect, the pump affects the signals directly. 
In particular, XPM and XPolM induces phase variations that impact the detection of phase modulated signals, and that are also converted into amplitude 
variations by group--velocity dispersion~\cite{Rapp-PTL-97}\cite{Rapp-JOC-99}. Since the distortions go back to random power fluctuations of the pump, 
they manifest itself as additional noise and are therefore difficult to remove. 

Since signal phase is not taken into account when using direct detection of amplitude modulated signals, there is no noticeable effect on transmission 
performance for on--off--keying (OOK) based signals. Thus, codirectional third--order Raman pumping is a suitable option for improving performance of 
such signals. However, the phase variations have a strong impact on the detection of signals making use of advanced modulation formats such as 
quaternary phase shift keying (QPSK). Therefore, higher--order codirectional pumping does not provide a noticeable benefit as compared with first order 
pumping. 

In this paper, a technique for reducing the impact of the described phase variations on symbol detection is presented. Reducing the impact 
of disturbing phase variations allows to use higher order pumping schemes for codirectional Raman amplification in order to improve the 
optical signal--to--noise ratio of existing fiber links and thus to increase their capacity. For this purposes, the presented
compensation technique makes use of the correlation of phase variations induced via XPM in neighboring signals copropagating in an 
optical fiber and resulting from intensity fluctuations of the fundamental pump.

\section{Distortions of the constellation diagram}
Higher--order pumping transfers optical power from a high--power pump to the signals via intermediate lightwaves called seeds. This is accomplished by 
taking advantage of the limited bandwidth of SRS with a peak of the Raman gain curve at a frequency shift of around 13\THz~\cite{Agrawal-NLFO-07}. When 
doubling the frequency shift, the Raman gain coefficient equals to less than $6\%$ of its peak value. The wavelengths of the pumps and the seeds are 
chosen such that the spacing between neighboring lightwaves is in the order of magnitude of the frequency shift of the peak. Thus, power transfer via SRS 
is only possible from one lightwave to its neighboring lightwave. 

Within the first kilometers of the transmission fiber, the power of the pump is transferred 
to the first seed without amplifying the signals significantly. In a next step, the power is transferred from the first seed to the second seed. Finally, 
this second seed provides amplification to the signals which reach their total power maximum after a propagation distance of approximately~$40\km$. 
Amplification of the signals occurs mainly within this region. However, this also is where significant fluctuations of the signal power levels 
can be induced due to the RIN transfer. The large power of the pump within the first kilometers of the transmission fiber is the root cause for 
above mentioned phase variations deteriorating system performance. 

The effect of strong phase noise on symbol detection is illustrated in Fig.~\ref{figure_phase_distortions} showing the constellation diagram including decision 
regions. On the left side, the constellation diagram is affected by additive noise only, whereas there is also phase noise on the right side. 

\graficeps{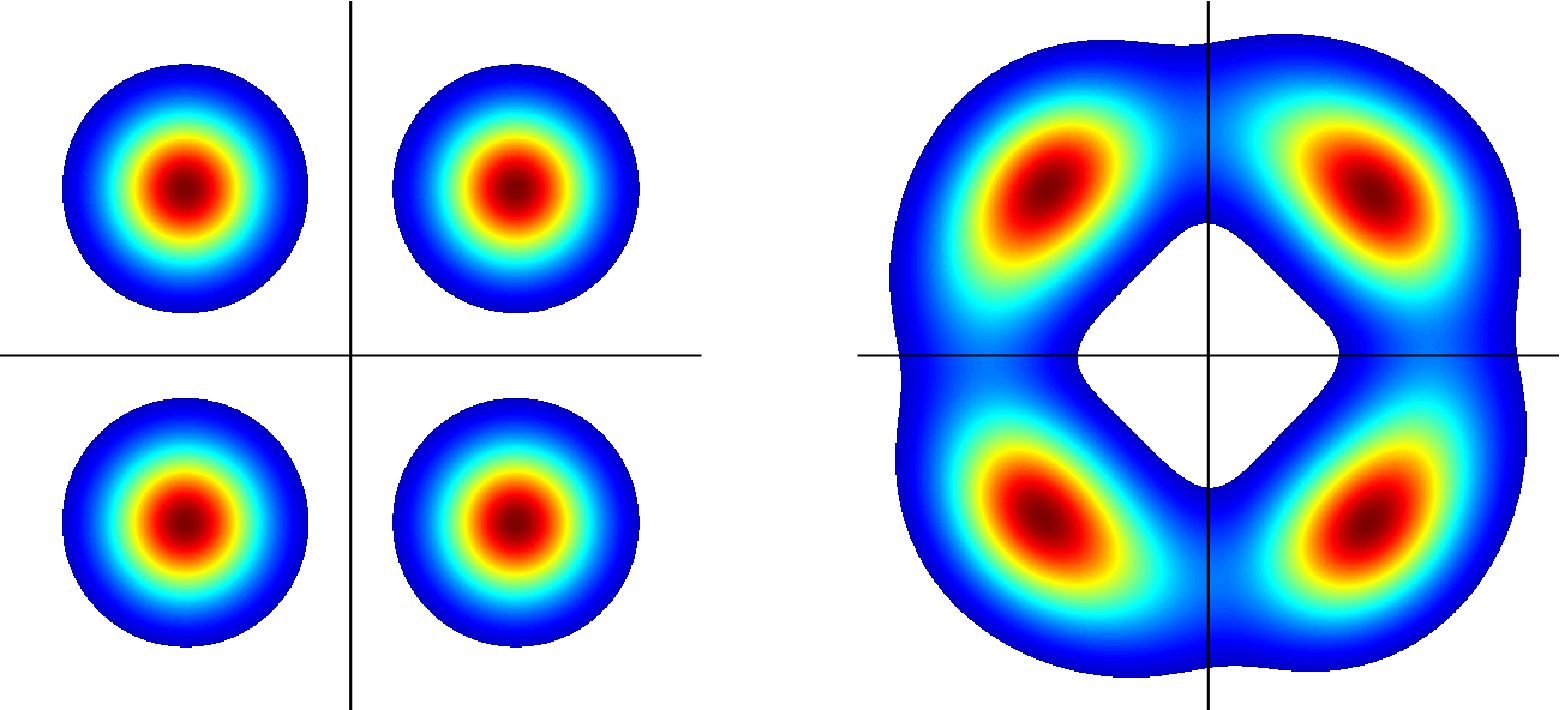}{0.8\figwidth}{Constellation diagram with phase noise only (left side) and in the presence of phase noise 
and additive noise (right side)}{htb}

Frequency characteristics of the conversion process is studied by assuming sinusoidal modulation of the power
of the pump signal with varying modulation frequency. Neglecting signal distortions by linear and nonlinear effects,
the efficiency of the process converting power variations of the pump into phase variations of a copropagating 
signal can mathematically be described by the equation
\begin{equation}
\eta \propto \sqrt{\alphaNp^2 + \left(\dphaseconst^{(s1p)}\cdot\omega\right)^2}
\end{equation}
wherein $\alphaNp$ stands for the attenuation coefficient of the pump
\begin{equation}
\alphaNp = \ln(10)/10\cdot\alphadB
\end{equation}
with $\alphadB$ being the attenuation coefficient in decibel units. Furthermore, $\omega$ stands
for the angular frequency and $\dphaseconst^{(s1p)}$ represents the difference between the inverses
of the group velocities 
\begin{equation}
\dphaseconst^{(s1p)} = \phaseconst^{(s1)} - \phaseconst^{(p)}
\end{equation}
of the pump and the signal. The resulting frequency characteristics is illustrated in Fig.~\ref{figure_efficiency}. Please note that no phase estimation removing 
slow phase variations is assumed for this representation. Simulation results show that this equation describes the frequency dependence of the conversion effect 
including linear and nonlinear effects quite well. 

\graficeps{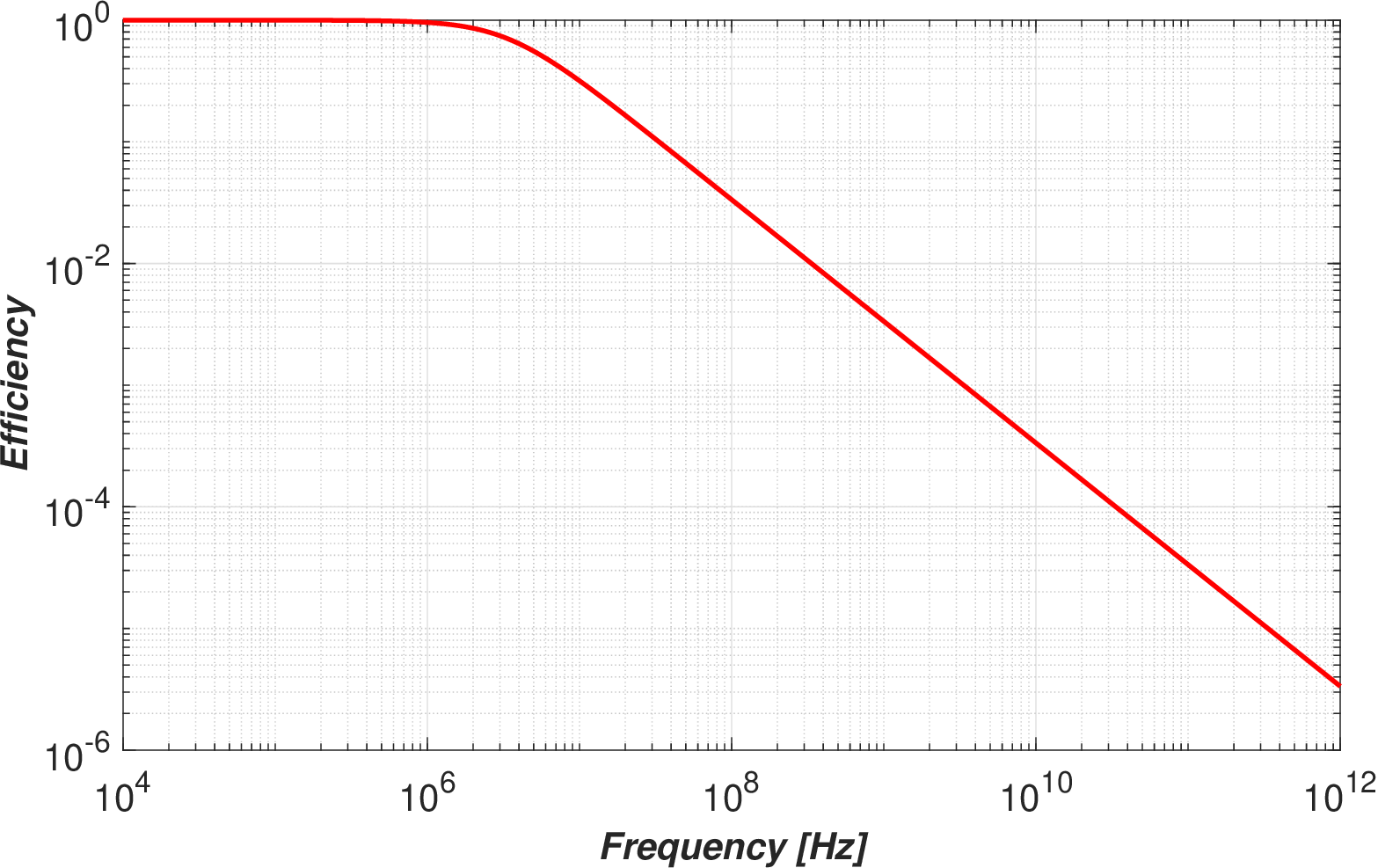}{0.8\figwidth}{Efficiency of power to phase conversion via XPM versus frequency}{htb}

In real systems, slow phase variations up to frequencies of around $1\MHz$ are already eliminated by carrier phase estimation performed by the digital signal 
processor (DSP) and random phase variations are observed at higher frequencies only. With increasing frequency, the magnitude passes through a maximum and 
finally diminishes continuously. This decrease is due to the walk--off effect between the pump signal and the affected signals. 

Two signals propagating in the same direction in an optical fiber are affected in a very similar way by XPM induced by intensity fluctuations of a strong 
pump as long as the walk--off effect between the two channels is small enough. Thus, the phase variations of copropagating signals are correlated 
which allows for reducing the impact on BER.

Neglecting fiber attenuation and other linear fiber effects, the complex envelope of the electrical fields of the two signals at the receivers 
$\signalRx^{(1)}(t)$ and $\signalRx^{(2)}(t)$ are represented by the following equations, wherein 
$\signalTx^{(1)}(t)$ and $\signalTx^{(2)}(t)$ stand for the respective complex envelopes at the transmitter.

\begin{eqnarray}
\signalRx^{(1)}(t) & = & \signalTx^{(1)}(t)\cdot{}\underbrace{e^{\imath\phasecommon}}_{\parbox{1.8cm}{\centering Common phase shift}} + \underbrace{\noise_{I}^{(1)} 
+ \imath \noise_{Q}^{(1)}}_{\mbox{Noise signal \rom1}} \\[1em]
\signalRx^{(2)}(t) & = & \signalTx^{(2)}(t)\cdot{}\underbrace{e^{\imath\phasecommon}}_{\parbox{1.8cm}{\centering Common phase shift}} + \underbrace{\noise_{I}^{(2)} 
+ \imath \noise_{Q}^{(2)}}_{\mbox{Noise signal \rom2}}
\end{eqnarray}

Both signals suffer from the same phase shift $\phasecommon$ induced by XPM due to intensity variations of the 
Raman pump. This phase shift is modeled as a Gaussian distributed random variable. Furthermore, additive noise represented by 
$\noise_{I}^{(1)} + \imath \noise_{Q}^{(1)}$ and $\noise_{I}^{(2)} + \imath \noise_{Q}^{(2)}$ is added
to each of the signals. Each of the variables $\noise_{I}^{(1)}$, $\noise_{Q}^{(1)}$, $\noise_{I}^{(2)}$
and $\noise_{2}^{(1)}$ represents stochastically independent random variables with Gaussian probability 
distribution. 

The key idea is to determine from the actual phase of both signals an estimate of the common phase shift.
In a next step, this common phase shift is removed from the signals: 

\begin{eqnarray}
\signaldet^{(1)}(t) & = & \signalRx^{(1)}(t)\cdot{}e^{-\imath\phasecommonest} \nonumber \\
                    & = & \left\{\signalTx^{(1)}(t)\cdot{}e^{\imath\phasecommon} + \noise_{I}^{(1)} 
                              + \imath \noise_{Q}^{(1)}\right\}\cdot{}e^{-\imath\phasecommonest} \\
								    & = & \signalTx^{(1)}(t)\cdot{}e^{\imath(\phasecommon-\phasecommonest)} 
												      + \left\{\noise_{I}^{(1)} 
                              + \imath \noise_{Q}^{(1)}\right\}\cdot{}e^{-\imath\phasecommonest}\\[1em]
\signaldet^{(2)}(t) & = & \signalRx^{(2)}(t)\cdot{}e^{-\imath\phasecommonest} \nonumber \\
                    & = & \left\{\signalTx^{(2)}(t)\cdot{}e^{\imath\phasecommon} + \noise_{I}^{(2)} 
                              + \imath \noise_{Q}^{(2)}\right\}\cdot{}e^{-\imath\phasecommonest} \\
									  & = & \signalTx^{(2)}(t)\cdot{}e^{\imath(\phasecommon-\phasecommonest)} 
												      + \left\{\noise_{I}^{(2)} 
                              + \imath \noise_{Q}^{(2)}\right\}\cdot{}e^{-\imath\phasecommonest}
\end{eqnarray}

\section{Compensation of the direct impact of the fundamental pump}
In the following, a very simple approach for determining the estimate of the common phase is presented.
For sure, this is in no way limiting. Determining the phases $\phaseactual^{(1)}$ and $\phaseactual^{(2)}$ 
of the two received signals by using the Viterbi--Viterbi phase estimator, the estimate of the common phase 
is calculated by using the equation
\begin{equation}
\phasecommonest = \frac{\weightfactor^{(1)}\cdot\phaseactual^{(1)} + \weightfactor^{(2)}\cdot\phaseactual^{(2)}}{\weightfactor^{(1)}+\weightfactor^{(2)}}
\end{equation}
with
\begin{eqnarray}
\weightfactor^{(1)} = \exp\left( -\probfactor{}\cdot\left|\phaseactual^{(1)}\right|\right) \\[1em]
\weightfactor^{(2)} = \exp\left( -\probfactor{}\cdot\left|\phaseactual^{(2)}\right|\right) 
\end{eqnarray}
and $\probfactor$ being a factor equal to or larger than zero. In the border case of very large values of the parameter
$\probfactor$, the estimate of the common phase corresponds to the phase having minimum magnitude. Depending on the symbol definition , 
$\sfrac{\pi}{2}$ needs to be subtracted from $\phasecommonest$.   

Several constellations are considered in the following for illustrating the functioning of the compensation
algorithm. For illustration purposes, it is assumed that both signals transmit the symbol found in the 
light blue quadrant of the constellation diagram and the above mentioned border case is assumed. In each of the 
plots, the common phase shift is illustrated by a red arc, whereas the signal individual additive noise if 
represented by a blue arrow. The position of the samples after common phase compensation is indicated 
by green dots. 

\graficeps{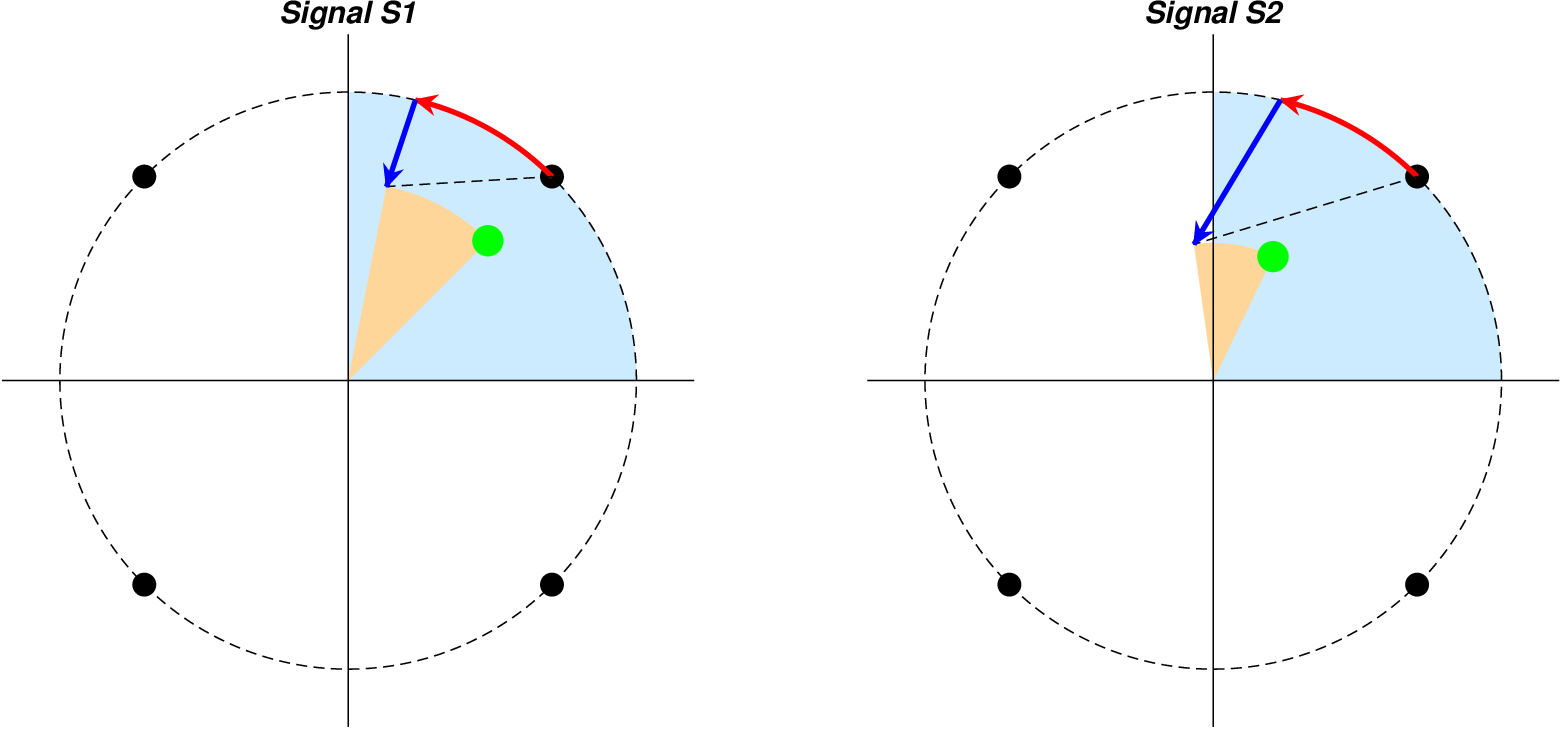}{0.7\figwidth}{Illustration of the compensation algorithm for 
the case wherein the symbol of the first signal ${\rm S}_1$ is found in the original quadrant, whereas
the symbol of the second signal ${\rm S}_2$ is moved to the neighboring quadrant due to additive
noise.}{htb}

\graficeps{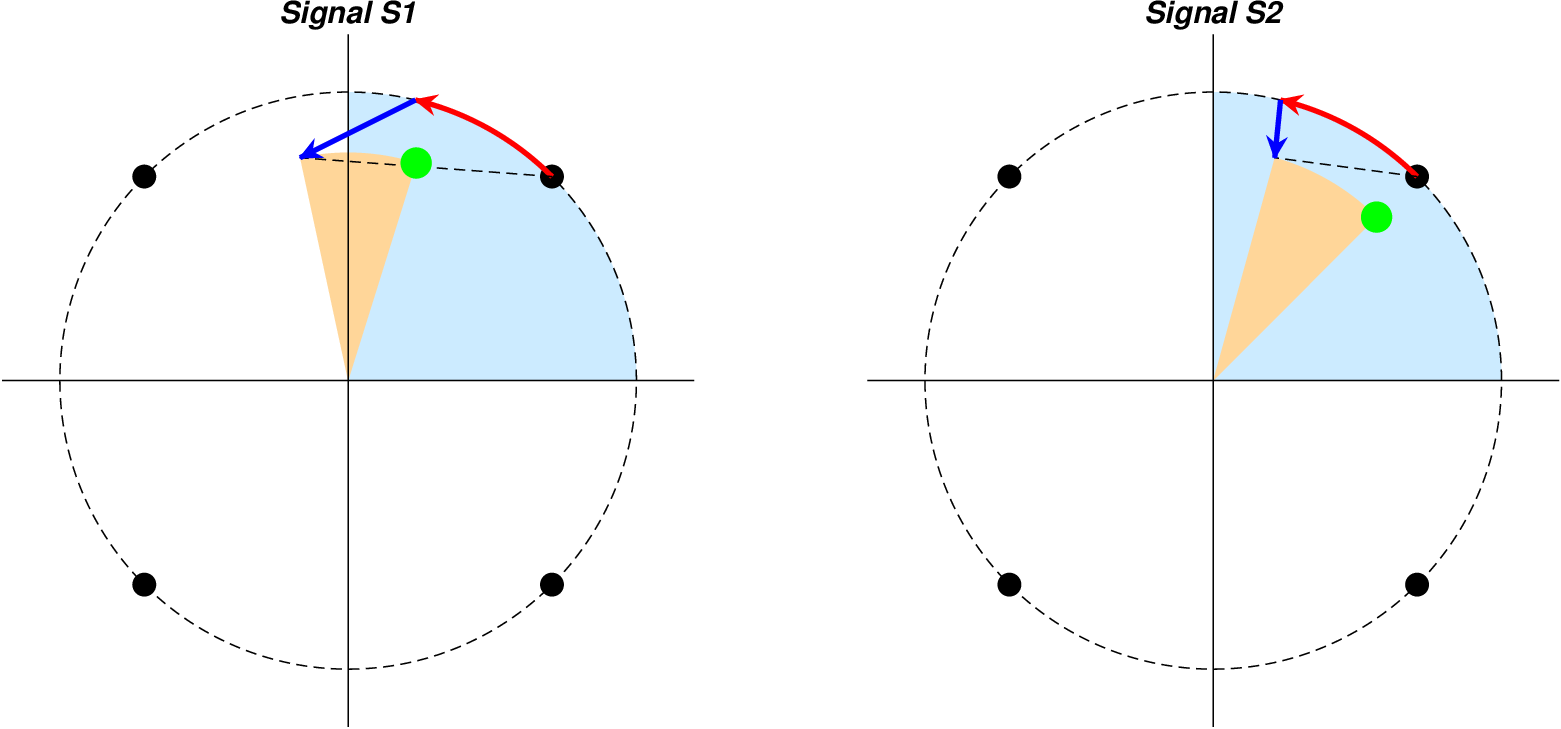}{0.7\figwidth}{Illustration of the compensation algorithm for 
the case wherein the symbol of the second signal ${\rm S}_2$ is found in the original quadrant, whereas
the symbol of the first signal ${\rm S}_1$ is moved to the neighboring quadrant due to additive
noise.}{htb}

In Figs.~\ref{figure_correction_2} and~\ref{figure_correction_4}, one of the symbols is moved to 
a neighboring quadrant, whereas the second symbol is found in the original quadrant. In both cases,
the compensation algorithm is able to move the symbol found in the wrong quadrant back into the 
original quadrant such that error free detection becomes possible since the magnitude of the phase of the 
symbol moved into the neighboring quadrant is larger as compared with the symbol in the original 
quadrant. 

\graficeps{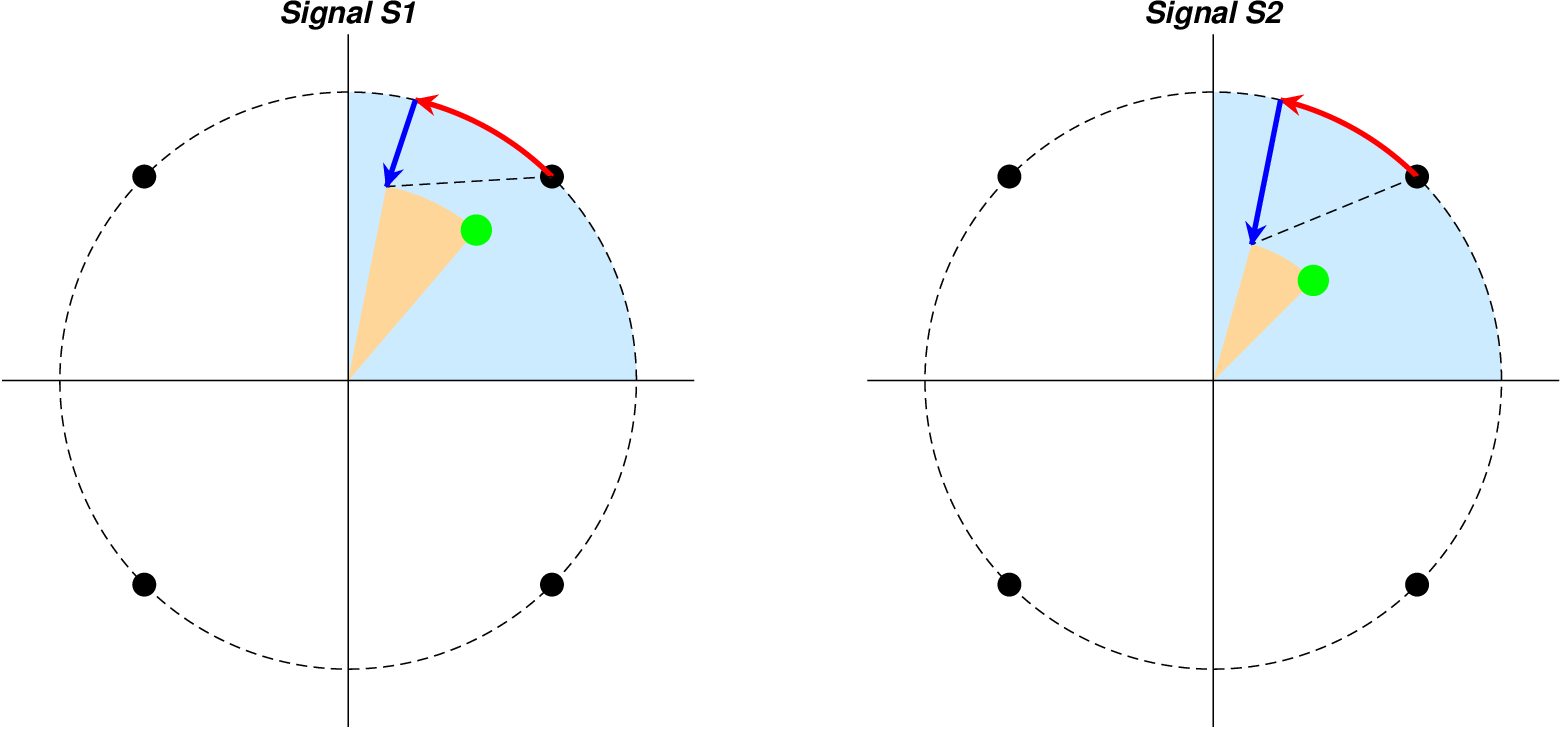}{0.7\figwidth}{Illustration of the compensation algorithm for 
the case wherein both received symbols are found in the original quadrant.}{htb}

As illustrated in Fig.~\ref{figure_correction_1}, the algorithm does not introduce additional 
errors in case both symbols are found in the original quadrant and are therefore mapped to the 
correct symbols even without common phase compensation. Furthermore, Fig.~\ref{figure_correction_3}
shows that the algorithm cannot reduce the bit error ratio in case both symbols are moved away
from the original quadrant.  

\graficeps{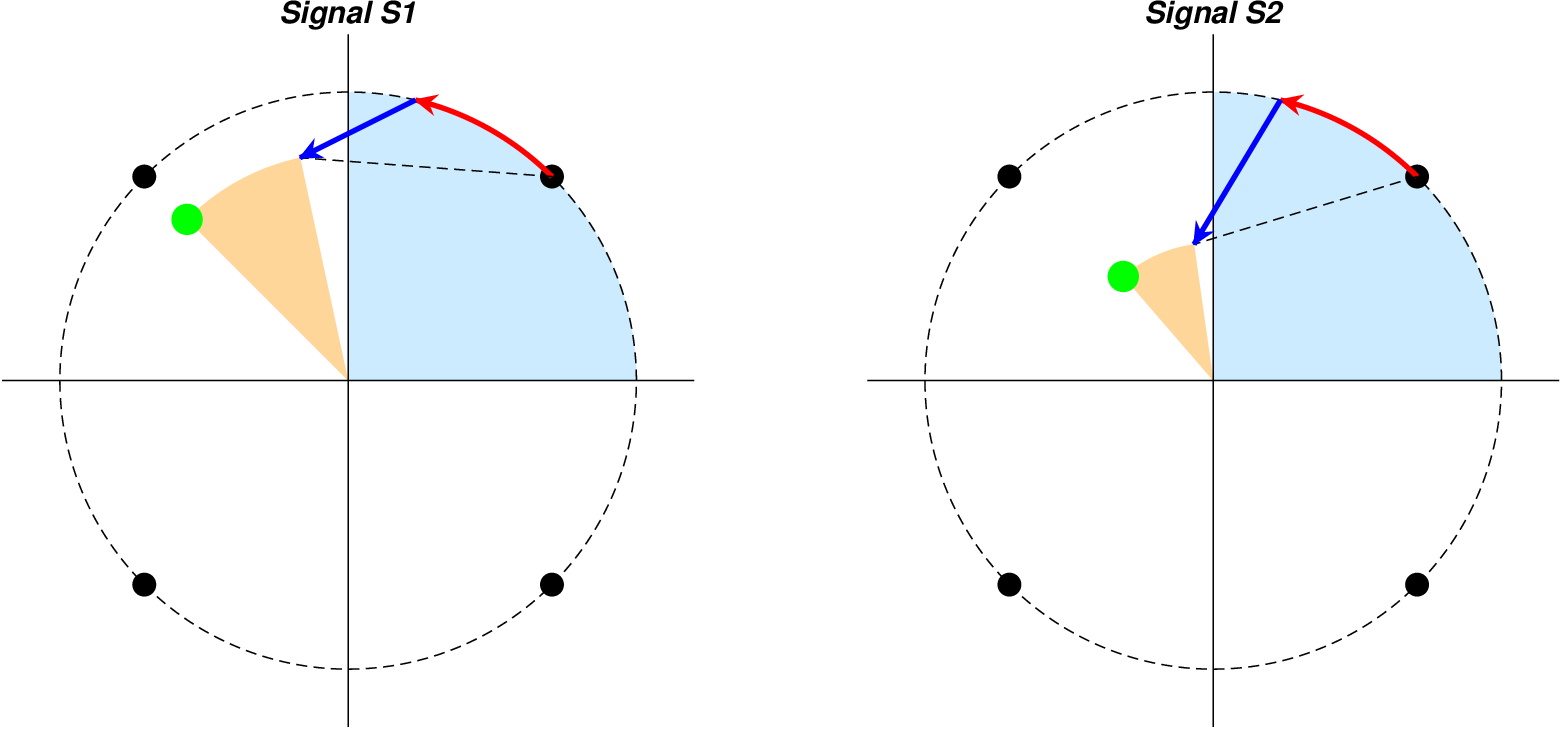}{0.7\figwidth}{Illustration of the compensation algorithm for 
the case wherein the symbol of both signals are moved to the neighboring quadrant due to additive
noise.}{htb}

In some cases, the compensation algorithm might even induce additional bit errors, as illustrated
in Fig~\ref{figure_correction_5}. However, the probability for this constellation is quite small 
as compared with the constellations wherein the compensation algorithm improves performance.
Therefore, the compensation algorithm is able to provide an overall performance improvement. 

\graficeps{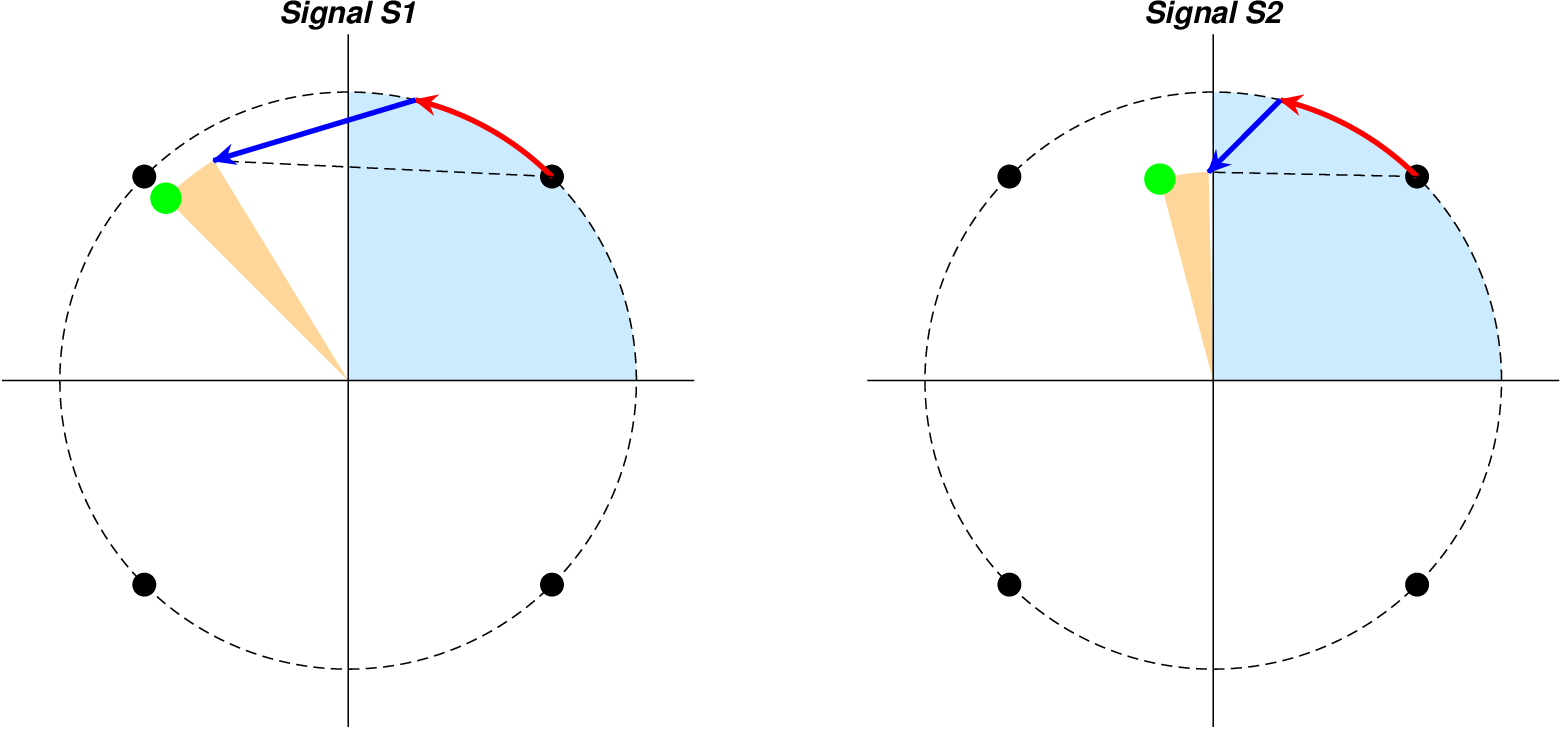}{0.7\figwidth}{Illustration of a constellation wherein the compensation 
algorithm induces additional bit errors.}{htb}

In summary, the following constellations are relevant:

\begin{center}
\begin{tabular}{|l|l|l|}
\hline
\multicolumn{3}{|c|}{Relevant constellations}\\
\hline
\hline
(1) & No correction required & Fig.~\ref{figure_correction_1} \\
\hline
(2) & Correction successful & Figs.~\ref{figure_correction_2} and~\ref{figure_correction_4}  \\
\hline
(3) & Additional errors & Fig.~\ref{figure_correction_5} \\
\hline
(4) & No correction possible & Fig.~\ref{figure_correction_3} \\
\hline
\end{tabular}
\end{center}

Although the algorithm may introduce some additional errors in some constellations, an overall reduction of the bit error ratio is 
achieved since constellations for which a successful correction is possible are more likely to happen. In other words, the probability
for a symbol shifted to the next quadrant having a smaller magnitude of the phase as compared with the symbol located in the 
original quadrant is quite small. 

\clearpage

\subsection{Processing steps}
Steps required for performing the described compensation of common phase variations are listed in the 
following. Some of them are illustrated in Fig.~\ref{figure_explanation}. 

\begin{enumerate}
\item Common processing of samples from both signals for reducing common phase variations.
\item At the transmitters, the clocks of the copropagating data signals need to be adjusted relative to each other in 
such a way that the phase variations at the later sampling points are affected by the same power
fluctuations of the pump. Essentially, this will result in synchronous clock signals, but some 
slight deviations might even provide better performance. The optimum time shift between the clock 
signals can be determined by means of a feedback control comprising the transmitter and the 
receiver. 
\item The two signals will arrive at the receiver with same delay. Thus, the data samples of the 
"faster" channels need to be buffered. Furthermore, compensation at higher frequencies requires
to remove some phase shift in the frequency domain.
\item The parameter $\probfactor$ determining the weight of the different contributions is adjusted
in another control loop. 
\end{enumerate}

\graficeps{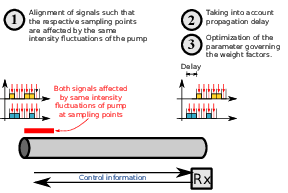}{0.9\figwidth}{Processing steps required for compensating common 
phase variations}{htb}

\clearpage
\section{Implementation aspects}
Using the Viterbi \& Viterbi phase recovery algorithm, data dependency is removed from the samples by raising them 
to the fourth power and calculating the phase of the results. Afterwards, the influence of additive noise is 
minimized by filtering. The algorithm is illustrated in Fig.~\ref{figure_Viterbi_Viterbi}.

\graficeps{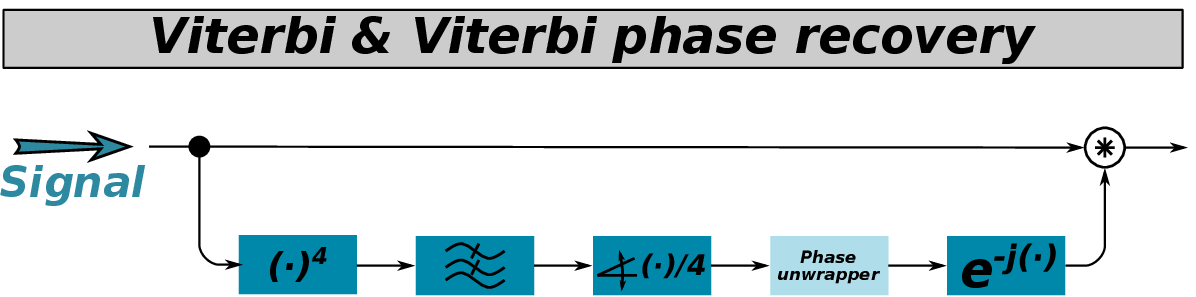}{0.8\figwidth}{Block diagram illustrating the functionality of the 
Viterbi \& Viterbi carrier phase estimation (CPE) algorithm}{htb}

The symbols of a quaternary phase shift keying (QPSK) signal can be described by the equation
\begin{eqnarray}
s_k & = & \exp\left\{ \imath k \cdot\frac{\pi}{2} + \varphi(t)\right\} \\
    & = & \exp\left\{ \imath k \cdot\frac{\pi}{2} + \bar{\varphi} + \tilde{\varphi}(t)\right\}		
\label{Eq:symbols}
\end{eqnarray}
with an arbitrary time dependent phase $\varphi(t)$ with average $\bar{\varphi}$ and 
fluctuations $\tilde{\varphi}(t)$ with zero average. 
The characteristic mathematical operation of the Viterbi \& Viterbi phase recovery algorithm is given by the 
equation
\begin{equation}
\phi(t) = \frac{1}{4}\cdot\angle\left\{ \left(s_k^\star\right)^4 \right\}
\end{equation}
wherein $s_k^\star$ stands for the detected samples and $\angle$ denotes the phase of the following 
expression in brackets (argument). With equation~(\ref{Eq:symbols}), this leads to the expression
\begin{equation}
\phi(t) = \bar{\varphi} + \tilde{\varphi}(t)
\end{equation}
The average phase $\bar{\varphi}$ is canceled out by carrier phase estimation, e.\,g. by using the 
Viterbi \& Viterbi approach. What remains are fast variations $\tilde{\varphi}(t)$ impairing symbol
detection. 

The constellation diagram at the receiver is rotated against the original constellation diagram at 
the transmitter if the phase of the local oscillator does not match the carrier phase. This mismatch
is canceled in the digital domain by carrier phase estimation (CPE). The fundamental functionality of 
the CPE is explained in the following starting from constellations rotated by various angles against
the original constellation diagram, as illustrated in Fig.~\ref{figure_start_phase}. 

\graficeps{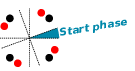}{0.5\figwidth}{Constellation diagram rotated against the original 
constellation diagram due to carrier phase mismatch}{htb}

Two possible implementations of the phase noise compensation algorithm in combination with 
Viterbi \& Viterbi phase recovery are illustrated in Fig.~\ref{figure_implementations}. 

\graficeps{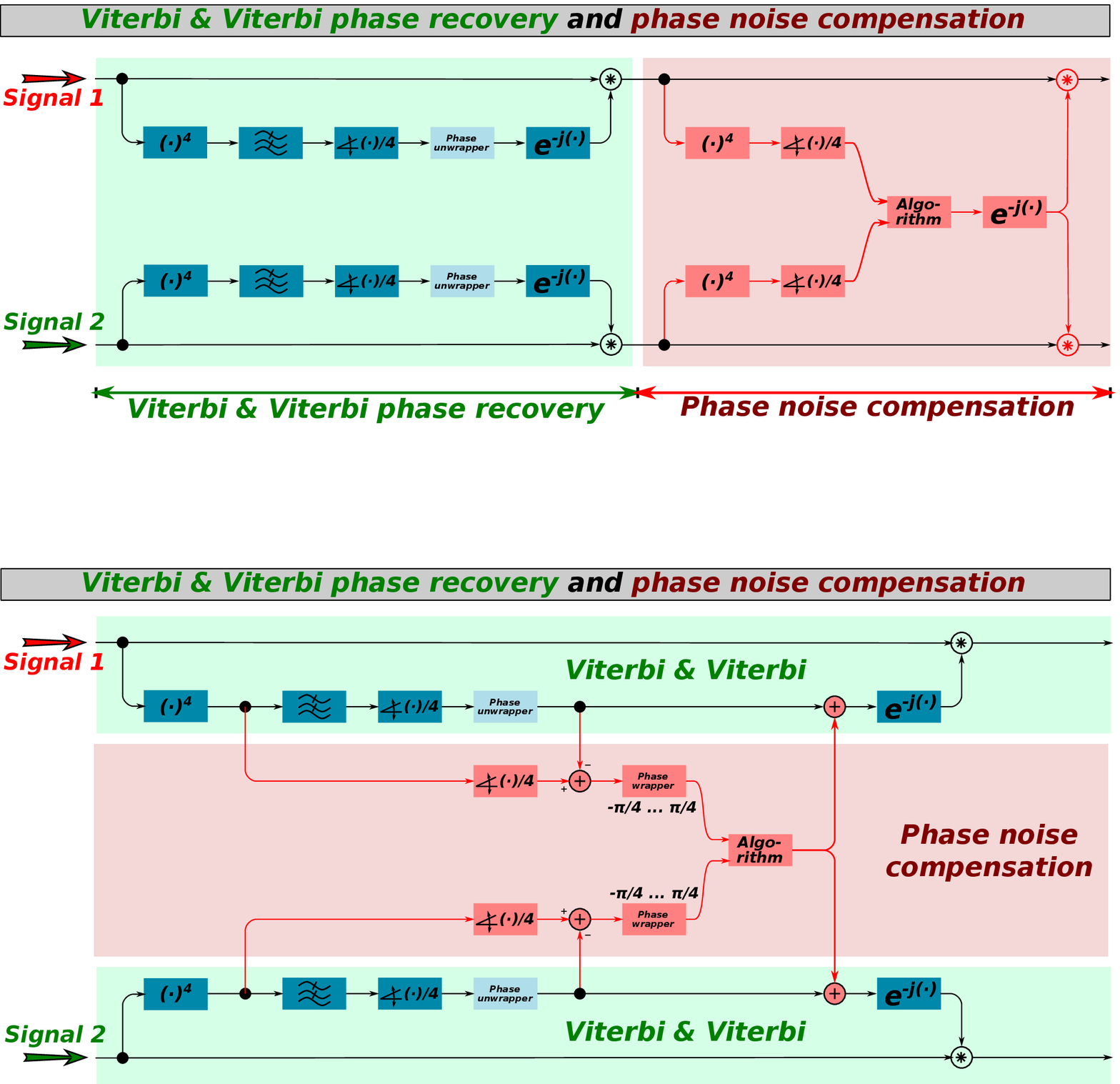}{\figwidth}{Two possible implementations of phase noise compensation 
in combination with Viterbi \& Viterbi phase recovery. In the upper diagram, carrier phase estimation 
and the inventive phase noise compensation are cascaded. In the lower diagram, both steps are 
combined in a single algorithm.}{htb}

\section{Conclusions}
Higher--order pumping schemes are suitable techniques for enhancing the performance of Raman amplifiers. Due to their complexity,
the main application area is to be found in unrepeatered submarine links, where often no cost--efficient alternatives are 
available. However, the increased pump powers present new challenges and give rise to additional noise contributions. 
Currently, quaternary phase--shift keying combined with polarization multiplexing and coherent detection (CP--QPSK) is the preferred 
modulation format for leading edge systems. Many installed unrepeatered submarine links are upgraded by introducing transponders
making use of this modulation format. 

Signal distortions of such phase--modulated signals induced directly by the high--power pump of a third--order codirectional Raman 
amplifier via the Kerr effect impair signal detection such that high--order codirectional pumping often does not provide any 
performance improvement as compared with first order pumping schemes. In contrast, there is clear improvement when using 
on--off--keying signals. In this paper, a technique reducing the impact of the detected phase distortions is presented in order
to make full use of the potential performance improvement expected from higher order pumping.

\end{document}

%% file: UNITS.TEX
\newcounter{roman1}
\newcommand{\rom}[1]{\setcounter{roman1}{#1}{\Roman{roman1}}}

\newcommand{\km}{{\mbox{\,km}}}

\newcommand{\THz}{{\,\mbox{THz}}}  
          
\newcommand{\MHz}{{\,\mbox{MHz}}}
